\documentclass[reprint,nofootinbib,superscriptaddress,amsmath,amssymb,aps,prd]{revtex4-2}

\usepackage{acronym}
\usepackage{amsfonts,amsmath,amssymb}
\usepackage{bm}
\usepackage{cases}
\usepackage{comment}
\usepackage[T1]{fontenc} 
\usepackage{graphicx}
\usepackage{indentfirst}
\usepackage{mathrsfs}
\usepackage{multirow}
\usepackage{scrextend}
\usepackage{xcolor}
\usepackage{soul}
\usepackage[most]{tcolorbox}
\usepackage{float}
\usepackage{ulem}
\usepackage{subcaption}
\usepackage{amsbsy,color,amsthm}
\usepackage{bbm}
\usepackage{times}
\usepackage{units}
\usepackage{dcolumn}
\usepackage[colorlinks,linkcolor=red,anchorcolor=green,citecolor=blue,CJKbookmarks=True]{hyperref}
\DeclareMathSymbol{\shortminus}{\mathbin}{AMSa}{"39}
\usepackage{braket}
\usepackage{txfonts}
\usepackage{enumitem}
\usepackage{orcidlink}

\newcommand{\meq}[1]{(\ref{#1})}

\begin{document}

\title{Gravitational-Bumblebee perturbations: Exact decoupling and isospectrality}
\author{Hui-Fa Liu\orcidlink{0000-0001-9567-1461}}
\email{hfaliu@163.com}
\thanks{The authors are listed in alphabetical order.}
\affiliation{Beijing University of Technology, Beijing 100124, China}

\author{Wentao Liu\orcidlink{0009-0008-9257-8155}}
\email[]{liuwentao@lzu.edu.cn} 
\affiliation{Lanzhou Center for Theoretical Physics, Key Laboratory of Theoretical Physics of Gansu Province, 
Key Laboratory of Quantum Theory and Applications of MoE,
Gansu Provincial Research Center for Basic Disciplines of Quantum Physics, 
Lanzhou University, Lanzhou 730000, China}
\affiliation{Institute of Theoretical Physics $\&$ Research Center of Gravitation,
Lanzhou University, Lanzhou 730000, China}

\author{Yu-Xiao Liu\orcidlink{0000-0002-4117-4176}}
\email[]{liuyx@lzu.edu.cn } 
\affiliation{Lanzhou Center for Theoretical Physics, Key Laboratory of Theoretical Physics of Gansu Province, 
Key Laboratory of Quantum Theory and Applications of MoE, Gansu Provincial Research Center for Basic Disciplines of Quantum Physics, 
Lanzhou University, Lanzhou 730000, China}
\affiliation{Institute of Theoretical Physics $\&$ Research Center of Gravitation,
Lanzhou University, Lanzhou 730000, China}

\author{Qi Su\orcidlink{0009-0007-0066-2302}}
\email{sqphysics@outlook.com}
\affiliation{Beijing University of Technology, Beijing 100124, China}

\author{Ding-fang Zeng\orcidlink{0000-0001-5430-0015}}
\email{dfzeng@bjut.edu.cn}
\affiliation{Beijing University of Technology, Beijing 100124, China}


\begin{abstract}

In this paper, we present the exact decoupling of the full metric and bumblebee field perturbations in a Schwarzschild-like background. 
The coupled system reduces to four decoupled master equations, revealing in each parity sector a Schwarzschild-like gravitational sector and a Lorentz-violating Maxwell-like vector sector.
While Lorentz violation modifies the propagation speed of the emergent vector modes, we demonstrate that the gravitational master modes exhibit a ``dynamical immunity'' to the non-minimal Lorentz-violating coupling, and that the odd- and even-parity perturbations remain strictly isospectral. 
Our work provides a rare example in which Lorentz-violating couplings reshape the field reconstruction while leaving the gravitational ringdown spectrum intact.
This mismatch in propagation speeds suggests a possible timing signature of bumblebee vector dynamics in black hole perturbations, offering a theoretical route to testing spontaneous Lorentz symmetry breaking in the era of multi-messenger astronomy.

\end{abstract}
\maketitle

\section{Introduction}

Lorentz symmetry breaking (LSB) provides an important window into possible modifications of gravity beyond general relativity \cite{Colladay:1998fq,Carroll:2001ws,Kostelecky:2008ts}. 
A simple effective realization, which can arise as a low-energy limit of string theory \cite{Kostelecky:1988zi}, is the bumblebee model \cite{Kostelecky:1988zi,Bluhm:2004ep,Bluhm:2007bd}, in which a vector field acquires a nonzero vacuum expectation value and spontaneously breaks Lorentz symmetry.
Over the past decade, a broad class of solutions has been obtained in this theory with nonminimal gravity-bumblebee coupling, including exact spherically symmetric solutions \cite{Casana:2017jkc,Maluf:2020kgf,Filho:2022yrk,Chen:2025ypx,Li:2025tcd,Zhu:2025fiy,Liu:2025oho,Liu:2024axg}, axisymmetric solutions \cite{Liu:2024axg,Poulis:2021nqh,AraujoFilho:2024ykw,Ovcharenko:2026rvj}, and numerical solutions \cite{Xu:2022frb,Ji:2024aeg,Bailey:2025oun,Panotopoulos:2024jtn,Liu:2025swi,Luo:2026oxw}. 
These developments have in turn motivated extensive investigations of the physical consequences of LSB \cite{Xu:2026zgd,Xu:2023xqh,Chen:2020qyp,Liang:2022hxd,Lai:2025nyo,Liu:2025bpp,Tang:2025eew,Liu:2024wpa,Zhang:2023wwk,Xu:2025jvk,Quan:2026awx,Xiao:2025flt,QiQi:2026zwp,Ou:2026ohc,Liu:2026khl,Sui:2026pyf,Cabral:2026ukd,Ahmed:2026fix,AraujoFilho:2026oqc}, ranging from classical tests to various dynamical and astrophysical phenomena.

Among these consequences, extracting observable signatures from LSB backgrounds is of particular importance, and gravitational waves provide perhaps the most promising observational channel for this purpose \cite{LIGOScientific:2016aoc,LISA:2022kgy,ET:2025xjr,Luo:2025ewp,Ruan:2020smc,Bian:2025ifp,Long:2023vph,Long:2024axi,Dong:2023bgt,Dong:2026xab,Jing:2025zhh,Xia:2026aty,Chakraborty:2026qru}.
 Compared with test field perturbations \cite{Deng:2025uvp}, gravitational perturbations are more directly relevant in this context, since they are tied to the ringdown signal of gravitational waves \cite{Berti:2025hly}.
In Einstein-Bumblebee theory, however, existing studies, even for the simplest spherically symmetric background, have often been restricted either to odd-parity perturbations \cite{Mai:2024lgk} or to purely gravitational perturbations with the bumblebee field frozen \cite{Liu:2022dcn,Li:2025itp}.
From the viewpoint of theoretical consistency, this is unsatisfactory. 
Since bumblebee black hole solutions are obtained from the coupled gravitational-bumblebee field equations \cite{Casana:2017jkc}, their linear perturbations should likewise be treated as a genuinely coupled problem, rather than by freezing one sector from the outset. 
Whether this coupled perturbation system can be rigorously decoupled, and whether the resulting odd- and even-parity gravitational modes remain isospectral after such a decoupling \cite{Cardoso:2019mqo,Li:2023ulk,Pani:2013ija}, are still open questions. 
These issues constitute the motivation of the present work.

The core challenge in resolving these issues is that the Regge-Wheeler-Zerilli decoupling strategy \cite{Regge:1957td,Zerilli:1970se,Zerilli:1970wzz}, originally developed for metric perturbations on Schwarzschild backgrounds, cannot be directly carried over to the present Lorentz-violating gravito-vector system. 
Since then, extending such decoupling methods to coupled gravito-vector systems has constituted a persistent and nontrivial line of inquiry in black hole perturbation theory \cite{Zerilli:1974ai,Moncrief:1974gw,Moncrief:1975sb,Chandrasekhar:1979iz,Kodama:2003kk,Pani:2013wsa,Langlois:2021aji,Kase:2023kvq,Guo:2022rms,Liu:2023uft,Hirano:2024pmk,Pope:2025jgz}.
A prototypical example is the Reissner-Nordstr\"om black hole, whose coupled gravito-electromagnetic perturbations were investigated by Zerilli and Moncrief in 1974 \cite{Zerilli:1974ai,Moncrief:1974gw,Moncrief:1975sb}; a fully decoupled formulation was clarified only about five years later by Chandrasekhar in 1979 \cite{Chandrasekhar:1979iz}.
The Einstein-Bumblebee system presents an analogous but distinct problem. 
Unlike the Einstein-Maxwell case, where the vector field is a $U(1)$ gauge field \cite{Bluhm:2015dna}, the bumblebee field is subject to a fixed-norm constraint and is nonminimally coupled to curvature, while the potential term removes the usual Maxwell-type gauge freedom \cite{Bailey:2006fd}.
These features prevent a direct import of the Einstein-Maxwell decoupling strategy \cite{Chandrasekhar}, requiring the coupled perturbation problem to be revisited from the full Einstein-Bumblebee field equations.

In this work, we revisit this problem for the Schwarzschild-like black hole in Einstein-Bumblebee theory \cite{Casana:2017jkc} by treating the metric and bumblebee field perturbations on the same footing.
Our aim is to determine whether the coupled gravitational-bumblebee system admits an exact master-equation formulation, despite the absence of the usual Maxwell-type gauge freedom.
We show that this is possible through a hierarchical decoupling procedure, in which the bumblebee sector is separated first and the induced gravitational sources are subsequently absorbed into suitable master variables.
This formulation allows us to identify the propagating degrees of freedom in both parity sectors and to examine the status of odd-even isospectrality in a controlled way.
The remainder of this paper is organized as follows. 
Section \ref{black-hole-solution} reviews the Einstein-Bumblebee theory. 
Section \ref{linear-perturbations} introduces linear perturbations and their harmonic decomposition. 
Section \ref{master-equations} derives the master equations governing the perturbation dynamics, while Sec. \ref{reconstruction} presents the reconstruction of the metric and bumblebee field perturbations.
Section \ref{quasinormal-modes} is devoted to the numerical calculation of QNMs. 
Section \ref{Discussion} concludes with a discussion of the main results and possible future directions. Throughout the paper, we adopt units $c=1$ and the metric signature $(-,+,+,+)$.


\section{Einstein-Bumblebee theory and background solution}
\label{black-hole-solution}

In Einstein-Bumblebee theory, a vector field $B_\mu$ acquires a nonzero vacuum expectation value, thereby inducing spontaneous local Lorentz symmetry breaking (LSB).
In the absence of additional matter fields, the action is
\begin{equation}
\begin{aligned}
\mathcal{S}=&\int d^{4}x\sqrt{-g} \bigg [  \frac{1}{2\kappa}\left(R-2\Lambda\right)+\frac{\xi}{2\kappa}B^{\mu}B^{\nu}R_{\mu\nu}
\\
&-\frac{1}{4} B_{\mu\nu}B^{\mu\nu}-V(B^{\mu}B_{\mu}\pm b^{2}) \bigg ]\,, 
\label{eq:action}
\end{aligned}
\end{equation}
where $\kappa=8\pi G$, $\xi$ is the nonminimal coupling constant with mass dimension $-2$, and $V(B^{\mu}B_{\mu}\pm b^{2})$ is the potential that fixes the norm of the bumblebee field to $\mp b^{2}$ in the vacuum.
The field strength tensor of the bumblebee field is defined as
\begin{align}
B_{\mu\nu}=\partial_{\mu}B_\nu-\partial_{\nu}B_\mu\,,
\end{align}
which is analogous to the Maxwell field strength tensor. 
A crucial distinction from standard electrodynamics is the presence of the potential $V(B^{\mu}B_{\mu}\pm b^{2})$, whose minimum fixes the norm of $B_{\mu}$ to a nonzero value.
The potential attains its minimum when
\begin{align}
B^{\mu}B_{\mu}=\mp b^{2}\,,
\label{eq:fixednorm}
\end{align}
where $b^{2}$ is a positive constant that sets the symmetry-breaking scale. 
The resulting nonzero vacuum expectation value of $B_{\mu}$ picks out a preferred direction in spacetime, thereby inducing LSB.

Variation of the action \eqref{eq:action} with respect to $g_{\mu\nu}$ and $B_{\mu}$ gives the Einstein-Bumblebee field equations,
\begin{align}
G_{\mu\nu}+\Lambda g_{\mu\nu}&=\kappa T^{B}_{\mu\nu}\,,
\label{eq:Einstein}
\\
\nabla_{\mu}B^{\mu\nu}+\frac{\xi}{\kappa}B_\mu R^{\mu\nu}&=2V'B^{\nu}\,.
\end{align}
Here $T^{B}_{\mu\nu}$ is the effective energy-momentum tensor of the bumblebee field, given by
\begin{equation}
\begin{aligned}
T^{B}_{\mu\nu}&=
\frac{\xi}{\kappa}\Big[\frac{1}{2}B^{\alpha}B^{\beta}R_{\alpha\beta}g_{\mu\nu}-B_{\mu}B^{\alpha}R_{\alpha\nu}-B_{\nu}B^{\alpha}R_{\alpha\mu}
\\
&+\frac{1}{2}\nabla_{\alpha}\nabla_{\mu}\left(B^{\alpha}B_{\nu}\right)+\frac{1}{2}\nabla_{\alpha}\nabla_{\nu}\left(B^{\alpha}B_{\mu}\right)
\\
&-\frac{1}{2}\nabla^{2}\left(B_{\mu}B_{\nu}\right)-\frac{1}{2}g_{\mu\nu}\nabla_{\alpha}\nabla_{\beta}\left(B^{\alpha}B^{\beta}\right)\Big]
\\
&+2V'B_{\mu}B_{\nu} +B_{\mu}^{\ \alpha}B_{\nu\alpha}
-\big(V+ \frac{1}{4}B_{\alpha\beta}B^{\alpha\beta}\big)g_{\mu\nu}\,.
\label{eq:TB}
\end{aligned}
\end{equation}
In this work, we set $\Lambda\!=\!0$ and restrict our attention to background configurations lying at the minimum of the potential, for which $V\!=\!0$ and $V'\!=\!0$.
Assuming a static, spherically symmetric spacetime and a purely radial background bumblebee field, the field equations admit the following Schwarzschild-like black hole solution \cite{Casana:2017jkc}:
\begin{equation}
\begin{aligned}
ds^2&=-f(r)dt^2+\frac{1+\ell}{f(r)}dr^2+r^2(d\theta^2+\sin^2\theta d\phi^2)\,,
\\
B_{\mu}&=(0\,,b_r(r)\,,0\,,0)\,,
\end{aligned}
\end{equation}
with
\begin{align}
f(r)=1-\frac{2M}{r}\,,  \quad\quad  b_r(r)=b\sqrt{(1+\ell)/f(r)}\,.
\end{align}
Here $M$ is the mass parameter, and the LSB parameter $\ell\!=\!\xi b^2$ characterizes the deviation from the Schwarzschild geometry induced by the nonminimal coupling between gravity and the bumblebee field.

In particular, the LSB parameter $\ell$ modifies the radial metric component and thus deforms the background away from the Schwarzschild spacetime. 
Consequently, the dynamics of perturbations on this background are generally different from those in the Schwarzschild case.
This motivates a study of linear perturbations around the above background. 
In the following sections, we analyze the coupled linear perturbations of the Einstein-Bumblebee system, derive the corresponding decoupled master equations, and examine how LSB affects the different propagating modes and their QNM spectra.

\section{Linear perturbations of the Einstein-Bumblebee system}\label{linear-perturbations}
We now turn to linear perturbations around the above background and expand the metric and the bumblebee field as
\begin{align}
\tilde{g}_{\mu\nu}=g_{\mu\nu}+h_{\mu\nu}\,,\quad \tilde{B}_{\mu}=B_{\mu}+\mathfrak{b}_{\mu}\,,
\label{pert}
\end{align}
where $h_{\mu\nu}=\delta g_{\mu\nu}$ and $\mathfrak{b}_{\mu}=\delta B_{\mu}$ denote the metric and bumblebee field perturbations, respectively. 
Substituting these expansions into the full field equations and retaining only terms linear in the perturbations, we obtain the linearized equations:
\begin{align}
\delta G_{\mu\nu}-\kappa \delta T^{B}_{\mu\nu}=0\,,
\label{eq:perEinstein}
\\
\delta(\nabla_{\mu} B^{\mu\nu})+\frac{\xi}{\kappa}\Big(\mathfrak{b}_{\mu} R^{\mu\nu}+B_{\mu}\delta R^{\mu\nu}\Big)=0\,.
\label{eq:perBumblebee}
\end{align}
Here all covariant derivatives and curvature tensors are defined with respect to the background metric $g_{\mu\nu}$, and the term $\delta(V'B^\nu)$ does not contribute at linear order. 
Indeed, since the fixed-norm condition is imposed exactly, the perturbations satisfy $\delta(B^\mu B_\mu)\!=\!0$, which implies $\delta V'\!=\!V''\delta(B^\mu B_\mu)\!=\!0$. 
Together with $V'\!=\!0$ on the background, this gives $\delta(V'B^\nu)\!=\!0$. 
The fixed-norm condition also yields the linearized constraint
\begin{align}
2B_{\mu}\mathfrak{b}^{\mu}-h_{\mu\nu}B^{\mu}B^{\nu}=0\,.
\label{consBH}
\end{align}
Treated as a physically reasonable branch, equivalent to a low-energy effective limit where massive norm fluctuations are frozen \cite{Bluhm:2008yt}, this relation plays a crucial role in reducing the number of independent perturbation functions and must be imposed consistently together with the linearized field equations.


The spherical symmetry of the background spacetime allows us to decompose the perturbations in terms of spherical harmonics, which naturally separate into odd- and even-parity modes. 
To simplify the analysis, we first note that the perturbation equations are independent of the azimuthal number $m$, and we may therefore set $m=0$ without loss of generality. 
We then adopt the Regge-Wheeler gauge to fix the gauge freedom associated with infinitesimal coordinate transformations \cite{Regge:1957td}. 
For a given multipole number $L$, the perturbations can be written as follows.

In the odd-parity sector,
\begin{align}
h^{\rm odd}_{\mu\nu}&=
\begin{pmatrix}
	0 & 0 & 0 &  {h_0}(t,r) \\
	0 & 0 & 0 & {h_1}(t,r) \\
	0 & 0 & 0 & 0 \\
	{h_0}(t,r) & {h_1}(t,r) & 0 & 0 \\
\end{pmatrix}\sin\theta \frac{\partial}{\partial\theta} P_{L}(\cos\theta)\,,
\label{oddpergen}
\\
\mathfrak{b}^{\rm odd}_{\mu}&=\Big(0\,,\,0\,,\,0\,,\,u_3(t,r)\sin\theta \frac{\partial}{\partial\theta}\Big)P_{L}(\cos\theta)\,,
\end{align}
while in the even-parity sector,
\begin{align}
&\begin{aligned}
h^{\rm even}_{\mu\nu}=&
\begin{pmatrix}
	f(r){H_0}(t,\!r)\!\!\! & {H_1}(t,\!r) & 0 & 0 \\
	{H_1}(t,\!r) & \frac{1+\ell}{f(r)} H_2(t,\!r)\!\!\!& 0 & 0 \\
	0 & 0 & r^2K(t,\!r)\!\!\! & 0 \\
	0 & 0 & 0 & r^2K(t,\!r)\sin^2\theta  \\
\end{pmatrix}\\
&\times P_{L}(\cos\theta)\,,
\end{aligned}\label{evenpergen}
\\
&\begin{aligned}
\mathfrak{b}^{\rm even}_{\mu}=\Big(u_{0}(t,r)\,,\,u_{1}(t,r)\,,\,u_{2}(t,r)\frac{\partial}{\partial\theta}\,,\,0\Big)P_{L}(\cos\theta)\,.
\label{evengena}
\end{aligned}
\end{align}
Here $P_{L}(\cos\theta)$ denotes the Legendre polynomial of degree $L$. 
An important difference from the Einstein-Maxwell system appears in the vector sector. 
In the Maxwell case, the $U(1)$ gauge symmetry provides an additional gauge freedom that can be used to eliminate one component of the vector field perturbation. 
In contrast, the bumblebee potential removes the $U(1)$ gauge symmetry, so no analogous gauge freedom is available for $\mathfrak{b}_{\mu}$.

After performing the spherical harmonic decomposition, we obtain a coupled system of partial differential equations in the coordinates $(t,r)$. 
These equations encode the interactions between the metric and bumblebee perturbations, but their coupled structure makes a direct analysis of the dynamics difficult. 
To extract physical information, in particular the QNM spectrum, it is therefore necessary to reduce the system to one or more decoupled master equations. 
These master equations typically take a Schr\"odinger-like form, with the corresponding master variables encoding the physical dynamical degrees of freedom. 
In the next section, we derive the master equations for the odd- and even-parity sectors separately.

\section{Master variables and equations}\label{master-equations}

In this section, we derive the master variables and the corresponding decoupled wave equations for the odd- and even-parity sectors. 
As will be shown below, although the metric and bumblebee perturbations are coupled at the level of the original field equations, the system can still be reorganized into a set of decoupled master equations with a transparent dynamical interpretation.

\subsection{Odd-parity perturbations}

We first focus on the odd-parity sector. 
The system of coupled differential equations for the metric perturbations $h_0(t,r)$, $h_1(t,r)$ and the bumblebee perturbation $u_3(t,r)$ is given by
\begin{align}
\begin{aligned}
h_0''+\dot{h}_1'&-\frac{2}{r}\dot{h}_1-\frac{\mu^2+2f(r)}{r^2f(r)}h_0
\\
&+\frac{\ell}{f(r)b_r(r)}\Big[\dot{u}_3'-\frac{1-5f(r)}{2rf(r)}\dot{u}_3\Big]=0\,,~~~~
\end{aligned}    \label{odd-Gt}
\\
\begin{aligned}
\dot{h}_0'-\ddot{h}_1-\frac{2}{r}\dot{h}_0-\frac{\mu^2f(r)}{r^2}h_1+\frac{\ell}{b_r}\Big[\frac{\mu^2}{r^2}u_3+\frac{\ddot{u}_3}{f(r)}\Big]&=0\,,
\end{aligned}    \label{odd-Gr}
\\
\begin{aligned}
f(r){h}_1'+f'(r)h_1-\frac{1}{f(r)}\dot{h}_0-\ell\,\Big[\frac{{u}_3}{b_r(r)}\Big]'&=0\,,
\end{aligned}\label{odd-Gphi}
\\
\begin{aligned}
\!{u}_3''-\frac{1\!+\!\ell}{f(r)^2}\ddot{u}_3+\tfrac{f'(r)}{f(r)}u_3'+\frac{\tfrac{\ell^2}{\kappa b^2}-(1\!+\!\ell)(\mu^2\!+\!2)}{r^2f(r)}u_3 &
\\
-\frac{\ell~b_r(r)}{2\kappa\,b^2f(r)}\Big[\dot{h}_0'-\ddot{h}_1-\frac{2}{r}\dot{h}_0-\frac{\mu^2f(r)}{r^2}h_1\Big]&=0\,.~~~~
\end{aligned}\label{odd-u}
\end{align}
Here and in what follows, we use the shorthand notation $u' \equiv \partial u(t,r)/\partial r$ and $\dot{u} \equiv \partial u(t,r)/\partial t$. 
We also use
\begin{align}
f'(r)\equiv\frac{df(r)}{dr}=\frac{1-f(r)}{r}\,,\quad\quad \mu^2\equiv L^2+L-2\,.
\end{align}
These equations show that the metric perturbations and the bumblebee perturbation are coupled through terms proportional to the parameter $\ell$, reflecting the nontrivial coupling induced by the background bumblebee field. 
Since the odd-parity sector involves only three perturbation functions, the derivation of master equations remains tractable.

It follows from Eqs. \eqref{odd-Gr} and \eqref{odd-u} that the metric perturbations can be completely eliminated, leaving a closed equation for $u_3(t,r)$:
\begin{align}
\left[\frac{\partial^2}{\partial r_*^2}-\frac{1}{c^2_{\text{eff}}}\frac{\partial^2}{\partial t^2}-V_0(r)\right] u_3(t,r)=0\,,
\label{eq:masterpsi0}
\end{align}
with the effective potential
\begin{align}
V_0(r)=\frac{1}{c^2_{\text{eff}}}\frac{\mu^2+2}{r^2}f(r)\,,
\end{align}
where the tortoise coordinate $r_*=r+2M\ln\left(\frac{r}{2M}-1\right)$, which is formally the same as in the Schwarzschild case, and we introduce an effective propagation speed for the bumblebee perturbation in the LSB background:
\begin{align}
c_{\text{eff}}=1/\sqrt{1+\ell-\tfrac{\ell^2}{2\kappa\,b^2}}\,.
\label{def:ct}
\end{align}
It is determined jointly by the vacuum expectation value of the bumblebee field and the coupling constant.
It is worth emphasizing that the deviation of $c_{\text{eff}}$ from unity already encodes the influence of the metric perturbations on the bumblebee mode. 
In this sense, the effective propagation speed obtained here is not the result of an isolated vector perturbation \cite{Liu:2024oeq}, but of the coupled Einstein-Bumblebee system. 
This can be seen directly from Eqs. \eqref{odd-Gt}-\eqref{odd-u}: if the metric perturbations are artificially set to zero, then either $\ell=0$, in which case Lorentz violation disappears, or $u_3=0$, so that the bumblebee perturbation itself vanishes.

A gravitational master equation can be derived for the metric sector. 
From Eqs. \eqref{odd-Gt}-\eqref{odd-Gphi}, we obtain a Regge-Wheeler-type equation for the metric sector, in which the bumblebee perturbation enters as a source term,
\begin{align}
\left[\frac{\partial^2}{\partial r_*^2}-\frac{\partial^2}{\partial t^2}-V_1(r)\right]\tilde{\Psi}_1(t,r)=\ell~ S(u_3,u_3',\dots)\,,
\label{eq:psi1s}
\end{align}
with the effective potential
\begin{align}\label{master-eq1}
V_1(r)=f(r)\frac{\mu^2-1+3f(r)}{r^2},
\end{align}
and $S(u_3,u_3',\dots)$ is constructed solely from the bumblebee perturbation.
Here, we define
\begin{align}
\tilde{\Psi}_1(t,r)=h_0(t,r)+\frac{r}{2}\Big[\dot{h}_1(t,r)-h'_0(t,r)\Big]\,,
\end{align}
to be the Cunningham-Price-Moncrief master variable \cite{Cunningham:1978zfa,Cunningham:1979px,Cunningham:1980cp}. 
The source term $S$ can be expressed as the same differential operator appearing on the left-hand side acting on a suitable combination of $u_3$. 
This source term can therefore be eliminated by an appropriate redefinition of the master variable.
We therefore redefine the master variable as
\begin{equation}
\begin{aligned}
\Psi_1(t,r)=\,\tilde{\Psi}_1(t,r)-\frac{\ell~r}{2f(r) b_r(r)}\dot{u}_3(t,r)\,,
\end{aligned}
\end{equation}
after which Eq.  \eqref{eq:psi1s} reduces to the homogeneous form
\begin{align}
\left[\frac{\partial^2}{\partial r_*^2}-\frac{\partial^2}{\partial t^2}-V_1(r)\right] \Psi_1(t,r)=0.
\end{align}
This equation has exactly the same form as the Regge-Wheeler equation in general relativity \cite{Regge:1957td}. 
Therefore, although LSB enters the definition of the master variable and hence affects the reconstruction of the original perturbations, it does not modify the propagation of this sector, which remains governed by the same effective potential as in the Schwarzschild case.

\subsection{Even-parity perturbations}
We now turn to the even-parity sector. 
The linearized fixed-norm constraint \eqref{consBH} implies
\begin{align}
u_1(t,r)=\frac{1}{2}b_r(r)H_2(t,r)\,.
\label{u1exp}
\end{align}
The linearized Einstein equations \eqref{eq:perEinstein} yield seven independent relations, while the linearized bumblebee equations \eqref{eq:perBumblebee} provide three additional ones. 
Together with Eq. \eqref{u1exp}, these form a closed system of eleven equations for the coupled even-parity perturbations.

In the original system, many terms proportional to $\ell$ appear mixed with the metric perturbations, which makes a direct application of standard decoupling procedures inconvenient \cite{Zerilli:1970wzz}. 
To facilitate the analysis, we first reorganize the equations into a form in which the bumblebee contributions appear only on the right-hand side as source terms. 
This is achieved by using Eq. \eqref{u1exp} to eliminate terms such as $\ell H_2$ whenever possible, and by taking suitable linear combinations of the field equations to remove the remaining metric-dependent $\ell$-terms. 
After rearranging the equations, the linearized even-parity gravitational field equations can be written as
\begin{align}
\begin{aligned}
H_0-H_2&=S_{A}\,,
\end{aligned}\label{eq:gA}
\\
\begin{aligned}
f{}\dot{K}'\!-\!\frac{1\!-\!3f}{2r}\dot{K}-\frac{f}{r}\dot{H}_2-\frac{\mu^2\!+\!2}{2r^2}f{}{H_1}&=S_{tr}\,,
\end{aligned}\label{eq:gtr}
\\
\begin{aligned}
f{}{H'_1}+f'{}{H_1}-\dot{H}_2-\dot{K}&=S_{t\theta}\,,
\end{aligned}\label{eq:gt}
\\
\begin{aligned}
f{}K'\!+\!\dot{H}_1\!-\!f~{H'_0}\!-\!\frac{1\!-\!3f}{2r}{H_0}\!-\!\frac{1\!+\!f}{2r}{H_2}&=S_{r\theta}\,,
\end{aligned}\label{eq:gr}
\\
\begin{aligned}
fK''\!+\!\frac{1\!+\!5f}{2r}K'\!-\!\frac{\mu^2}{2r^2}K\!-\!\frac{f}{r}H'_2-\!\frac{\mu^2\!+\!4}{2r^2}H_2&=S_{tt}\,,
\end{aligned}\label{eq:gtt}
\\
\begin{aligned}
-\frac{1}{f}\ddot{K}+\frac{1+f}{2r}K'-\frac{\mu^2}{2r^2}K-\frac{f}{r}{H'_0}&
\\
+\frac{2}{r}\dot{H}_1+\frac{\mu^2+2}{2 r^2}{H_0}-\frac{1}{r^2}{H_2}&=S_{rr}\,,~~\,
\end{aligned}\label{eq:grr}
\\
\begin{aligned}
\!\!\! f{}K''\!-\!\frac{1}{f}\ddot{K}\!+\!\frac{1\!+\!f}{r}K'+2\dot{H}'_1\!+\!\frac{1\!+\!f}{rf}\dot{H}_1\!-\!f{}{H''_0}&
\\
-\frac{1}{f}\ddot{H}_2-\frac{3\!-\!f}{2r}H'_0
\!-\!\frac{1\!+\!f}{2r}H'_2+\frac{\mu^2\!+\!2}{2r^2}\Big(H_0\!-\!H_2\Big)&=S_{B}\,,~~\,
\end{aligned}\label{eq:gB}
\end{align}
where the source terms on the right-hand side depend only on the bumblebee perturbations:
\begin{align}
S_{A}&=\frac{2\ell}{b_r(r)}\Big[\frac{f'(r)}{2f(r)}u_2+u'_2\Big]\,,
\label{eq:sA}
\\
S_{tr}&=\frac{\ell(\mu^2+2)}{2r^2b_r(r)}\Big(-u_0+\dot{u}_2\Big)\,,
\\
S_{t\theta}&=\frac{\ell}{b_r(r)}\Big[\frac{f'(r)}{2f(r)}u_0+u'_0-\frac{1-5f(r)}{2rf(r)}\dot{u}_2+\dot{u}'_2\Big]\,,
\\
S_{r\theta}&=\frac{\ell}{b_r(r)f(r)}\Big[\dot{u}_0+\frac{2f(r)}{r^2}u_2-\ddot{u}_2\Big]\,,
\\
S_{tt}&=\frac{\ell(\mu^2+2)}{r^2b_r(r)}\Big(\frac{1}{r}u_2+u'_2\Big)\,,
\\
S_{rr}&=\frac{\ell}{rf(r)b_r(r)}\Big\{2\dot{u}_0+\frac{\mu^2+2}{2r^2}\Big[1+f(r)\Big]u_2\Big\}\,,
\\
S_{B}&=\frac{\ell}{b_r(r)}\Big\{\frac{2}{f(r)}\Big(\frac{\dot{u}_0}{r}\!+\!\dot{u}'_0\Big)\!+\!\frac{\mu^2\!+\!2}{r^2}\Big[\frac{f'(r)}{2f(r)}u_2\!+\!u'_2\Big]\Big\}\,.
\label{eq:sB}
\end{align}
In the original formulation, the even-parity bumblebee equations contain explicit metric dependent source terms, as shown in Appendix \ref{AppA}. 
After using the gravitational equations and the fixed-norm constraint, these metric dependent contributions can be reorganized, and the linearized even-parity bumblebee equations can be rewritten as
\begin{align}
{u''_0}{\!}-{\!}\dot{u}'_1+\frac{2}{r}\Big({u'_0}{\!}-{\!}\dot{u}_1\Big)-\frac{c^{-2}_{\text{eff}}(\mu^2{\!}+{\!}2)}{r^2f(r)}\Big({u_0}{\!}-{\!}{\dot{u}_2}\Big)&=0\,,
\label{eq:B1}
\\
\dot{u}'_0-\ddot{u}_1-\frac{\mu^2+2}{r^2}f(r)\Big({u_1}-{u'_2}\Big)&=0\,,
\label{eq:B2}
\\
u''_2{\!}-{\!}{u'_1}{\!}-{\!}\frac{c^{-2}_{\text{eff}}}{f(r)^2}\Big(\ddot{u}_2{\!}-{\!}\dot{u}_0\Big){\!}+{\!}\frac{f'(r)}{f(r)}\Big(u'_2{\!}-{\!}{u_1}\Big)&=0\,.
\label{eq:B3}
\end{align}
The structure of Eqs. \eqref{eq:gA}-\eqref{eq:gB} is now close to that of the linearized Einstein equations in the Schwarzschild case, with the main new feature being the source terms built purely from the bumblebee perturbations. 
Likewise, Eqs. \eqref{eq:B1}-\eqref{eq:B3} resemble the corresponding Maxwell perturbation equations, except that the propagation speed is replaced by the effective speed $c_{\text{eff}}$ defined in Eq. \eqref{def:ct}. 
This Maxwell-like closed form, however, should not be interpreted as a test vector limit: the metric perturbations have not been discarded; their effects have been absorbed through the coupled field equations into $c_{\rm eff}$ and the reconstruction relations in Sec. \ref{reconstruction}.
These structural similarities suggest that a Zerilli-type master formulation should exist  \cite{Zerilli:1970se,Zerilli:1970wzz,Zerilli:1974ai}. 
However, because the gravitational and bumblebee perturbations remain coupled, the original Zerilli procedure is not sufficient to decouple the system, and we instead construct the master equations using the more general method developed in Refs. \cite{Lenzi:2021wpc,Liu:2025reu,Liu:2025ipr}.

After the above rearrangement, the coupled even-parity system naturally gives rise to two master equations.
One is most conveniently obtained from the subset \eqref{eq:B1}-\eqref{eq:B3}, which forms a closed system for the variables $u_0(t,r)$, $u_1(t,r)$, and $u_2(t,r)$. 
We therefore introduce the master variable
\begin{align}
\Psi_2(t,r)=r^2\Big[u'_0(t,r)-\dot{u}_1(t,r)\Big]\,,
\end{align}
which satisfies
\begin{align}
\left[\frac{\partial^2}{\partial r_*^2}-\frac{1}{c^2_{\text{eff}}}\frac{\partial^2}{\partial t^2}-V_2(r)\right] \Psi_2(t,r)=0\,,
\end{align}
with the effective potential
\begin{align}
V_2(r)=\frac{1}{c^2_{\text{eff}}}\frac{\mu^2+2}{r^2}f(r)\,.
\label{master-eq2}
\end{align}
We note that, although the master variables are different, the wave equation and the effective potential coincide exactly with those in the odd-parity sector. 
This implies that the bumblebee modes in the two parity sectors are strictly isospectral.

The gravitational master equation can be extracted from Eqs. \eqref{eq:gA}-\eqref{eq:gB}, supplemented by the source structure displayed in Eqs. \eqref{eq:sA}-\eqref{eq:sB}. 
To capture the gravitational sector in a Zerilli-type form, we define the master variable
\begin{equation}
\begin{aligned}
\Psi_3(t,r)=&\frac{rf(r)}{\lambda(r)}H_2(t,r)+\frac{r}{2}K(t,r)
\\
&-\frac{r^2f(r)}{\lambda(r)}K'(t,r)+\frac{\ell}{b_r(r)}\frac{\mu^2+2}{\lambda(r)}u_2(t,r)\,,
\end{aligned}
\end{equation}
with 
\begin{align}
\lambda(r)=\mu^2+3-3f(r)\,.
\end{align}
The corresponding master equation reads
\begin{align}
\left[\frac{\partial^2}{\partial r_*^2}-\frac{\partial^2}{\partial t^2}-V_3(r)\right] \Psi_3(t,r)=0\,,
\end{align}
where the effective potential is
\begin{align}
V_3(r)=f(r)\frac{2(\mu^2+3)\mu^4+\lambda(r)^3}{3r^2\lambda(r)^2}\,.
\label{master-eq3}
\end{align}
This equation takes the standard Zerilli form, whereas the bumblebee perturbation enters only through the definition of the master variable.
This also implies that the gravitational perturbations are isospectral. 
Although the effective potentials in Eqs. \meq{master-eq3} and \meq{master-eq1} are different in form, they are related by a Darboux transformation \cite{Glampedakis:2017rar}, as in the standard Regge-Wheeler-Zerilli correspondence, and hence share the same spectrum.
In this sense, the gravitational sector exhibits a dynamical immunity to the Lorentz-violating coupling. 
The immunity is not a decoupling of the original metric variables from the bumblebee perturbations; rather, it refers to the invariance of the gravitational master equations after the bumblebee-induced sources have been absorbed into the redefined master variables. 
Lorentz violation therefore affects the reconstruction of the original perturbations, but leaves the Regge-Wheeler-Zerilli propagation structure and the gravitational QNM spectrum unchanged.

At the end of this section, let us briefly comment on the structure of the master equations obtained above. 
In both parity sectors, the perturbations separate into two dynamical classes. 
The variables $u_3(t,r)$ and $\Psi_2(t,r)$ obey Maxwell-like equations, while $\Psi_1(t,r)$ and $\Psi_3(t,r)$ satisfy the Regge-Wheeler and Zerilli equations, respectively.
This shows that, at the level of the master equations, the coupled Einstein-Bumblebee system can still be organized into propagation equations of familiar form.
However, this does not mean that the physical role of every perturbation function is already transparent.
In particular, the function $u_2(t,r)$ in the even-parity sector requires further clarification. 
In the Einstein-Maxwell case, the analogous function is pure gauge and can be removed by the underlying $U(1)$ symmetry. 
In the present Einstein-Bumblebee system, by contrast, the bumblebee potential removes the $U(1)$ gauge freedom, so $u_2(t,r)$ no longer admits the same straightforward gauge interpretation. 
To clarify its physical role and its contribution to the reconstructed perturbations, it is therefore useful to reconstruct the perturbations explicitly. 
This will be carried out in the next section.

\section{Reconstruction of the perturbations}\label{reconstruction}

Having obtained the master equations, we now reconstruct the perturbations by expressing the metric and bumblebee variables in terms of the master variables. 
This makes explicit how the propagating degrees of freedom are encoded in the original perturbation fields, and in particular clarifies the role of quantities such as $u_2(t,r)$ in the even-parity sector.

In the odd-parity sector, the perturbations are completely reconstructed from the two master variables $u_3(t,r)$ and $\Psi_1(t,r)$. 
The bumblebee perturbation is directly represented by $u_3(t,r)$, while the metric perturbations take the form
\begin{align}
h_1(t,r)&=-\frac{2r}{\mu^2f(r)}\dot{\Psi}_1(t,r)+\frac{\ell}{f(r)b_r(r)}u_3(t,r)\,,
\\
h_0(t,r)&=-\frac{2f(r)}{\mu^2}\Big[\Psi_1(t,r)+r\Psi'_1(t,r)\Big]\,.
\end{align}
These expressions show that the odd-parity metric perturbations are linear combinations of the two propagating modes $\Psi_1(t,r)$ and $u_3(t,r)$. 
Therefore, once the two master equations are solved, the odd-parity perturbations are completely determined.

The reconstruction in the even-parity sector is more involved. 
The bumblebee perturbations can be expressed as
\begin{align}
u_1(t,r)&=\frac{1}{(\mu^2+2)f(r)}\dot{\Psi}_2(t,r)+u'_2(t,r)\,,
\\
u_0(t,r)&=\frac{c^2_{\text{eff}}f(r)}{\mu^2+2}\Psi'_2(t,r)+\dot{u}_2(t,r)\,,
\end{align}
while the metric perturbations can be reconstructed in terms of the master variables and $u_2(t,r)$ as
\begin{widetext}
\begin{align}
H_2(t,r)&=\frac{2}{b_r(r)}\Big[\frac{1}{(\mu^2+2)f(r)}\dot{\Psi}_2(t,r)+u'_2(t,r)\Big]\,,
\\
H_0(t,r)&=\frac{2}{b_r(r)}\Big[\frac{1}{(\mu^2+2)f(r)}\dot{\Psi}_2(t,r)+(1+\ell)u'_2(t,r)\Big]+\frac{\ell}{b_r(r)}\frac{f'(r)}{f(r)}u_2(t,r)\,,
\\
K(t,r)&=\frac{4}{\mu^2+2}\Big\{\Big[\frac{\mu^2(\mu^2+3)}{3r\lambda(r)}+\frac{\mu^2}{6r}+f'(r)\Big]\Psi_3(t,r)+f(r)\Psi'_3(t,r)\Big\}-\frac{2\ell}{rb_r(r)}u_2(t,r)\,,
\\
H_1(t,r)&=\frac{4}{\mu^2+2}\Big\{\Big[\frac{\mu^2}{\lambda(r)}-\frac{1-f(r)}{2f(r)}\Big]\dot{\Psi}_3(t,r)+r\dot{\Psi}'_3(t,r)+\frac{\ell\,c^2_{\text{eff}}}{4b_r(r)}\Psi'_2(t,r)\Big\}\,.
\end{align}
\end{widetext}
Here $u_2(t,r)$ enters the reconstruction as a constrained quantity, and only its radial derivative is determined by the master variables:
\begin{equation}
\begin{aligned}
u'_2(t,r)=&\frac{b^2}{(\mu^2+2)b_r(r)}\Bigg\{\Big[\frac{2\mu^2}{\lambda(r)}-\frac{1-f(r)}{f(r)}\Big]\Psi'_3(t,r)
\\
&+\frac{r}{f(r)^2}V_3(r)\Psi_3(t,r)+\frac{2r}{f(r)^2}\ddot{\Psi}_3(t,r)\Bigg\}
\\
&-\frac{1}{(1+\ell)(\mu^2+2)f(r)}\dot{\Psi}_2(t,r)\,.
\end{aligned}
\end{equation}
A consistency check against the full set of linearized field equations confirms that the system determines only $u_2'(t,r)$, rather than $u_2(t,r)$ itself. Accordingly,
\begin{align}
u_2(t,r)=\int^r u'_2(t,\tilde r)\,d\tilde r + C(t)\,,
\end{align}
where $C(t)$ is an integration function.
It does not correspond to a new propagating degree of freedom, but must be fixed by the boundary or regularity conditions imposed on the reconstructed perturbations.

In summary, the reconstruction shows that the master equations obtained in the previous section fully characterize the propagating content of the perturbations. 
The original metric and bumblebee variables can then be reconstructed explicitly from the corresponding master variables, together with the constrained quantity $u_2(t,r)$ in the even-parity sector.
In this way, the relation between the propagating modes and the full perturbation variables becomes completely explicit. 
With both the master equations and the reconstruction formulas at hand, we are now in a position to study the spectral properties of the perturbations. 
In the next section, we turn to the calculation of the QNMs associated with these master equations.

\section{Quasinormal modes of the perturbations}\label{quasinormal-modes}

As shown in the previous sections, the perturbation problem of the Einstein-Bumblebee system reduces to four master equations. 
The equations for $\Psi_1(t,r)$ and $\Psi_3(t,r)$  take the Regge-Wheeler and Zerilli forms, respectively, and therefore have the same QNM spectra as in the Schwarzschild case. 
We do not discuss them further here. 
Meanwhile, the master variables $u_3(t,r)$ and $\Psi_2(t,r)$ obey the same master equation, and the corresponding odd- and even-parity perturbations are therefore isospectral. 
It is thus sufficient to analyze the QNMs associated with the master equation \eqref{eq:masterpsi0} for $u_3(t,r)$. 
In particular, we focus on how the effective propagation speed $c_{\text{eff}}$ modifies the spectrum.

\begin{figure*}[t]
\centering
\includegraphics[width=0.98\linewidth]{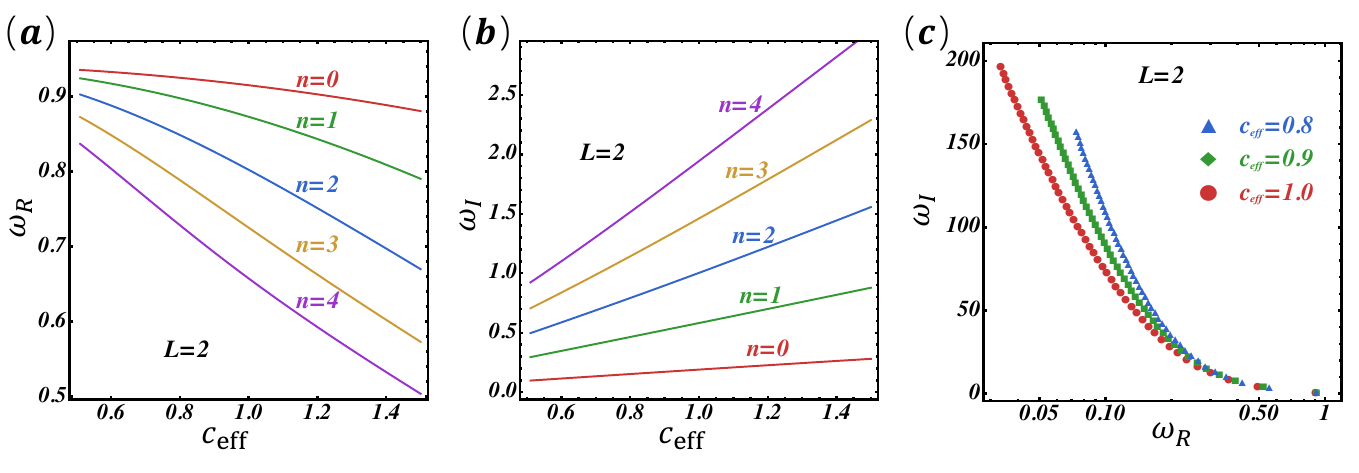}
\caption{QNM spectrum for $L=2$ at different values of $c_{\text{eff}}$. 
Panels (a) and (b) show the real and imaginary parts of the frequency for the first few overtones as functions of $c_{\text{eff}}$. 
Panel (c) displays the same spectral dependence in the complex-frequency plane for $c_{\text{eff}}=0.8$, $0.9$, and $1.0$, where every eighth overtone is shown up to $n=400$.} 
\label{fig1}
\end{figure*}
To cast Eq. \eqref{eq:masterpsi0} into the standard wave equation form, we introduce the rescaled tortoise coordinate
\begin{align}
x=\frac{1}{c_{\text{eff}}}\left[r+2M\ln\left(\frac{r}{2M}-1\right)\right]\,.
\end{align}
In terms of $x$, the equation becomes
\begin{align}
\left[\frac{\partial^2}{\partial x^2}-\frac{\partial^2}{\partial t^2}-c^2_\text{eff}V_0(r(x))\right]u_3(t,x)=0\,.
\end{align}
This form makes the QNM boundary conditions immediate:
\begin{align}
u_3(t,x)\sim e^{-i\omega(t+x)}\,, \qquad x\to -\infty\,,
\\
u_3(t,x)\sim e^{-i\omega(t-x)}\,, \qquad x\to +\infty\,.
\end{align}
These correspond to purely ingoing waves at the event horizon and purely outgoing waves at spatial infinity, and lead to a discrete set of complex frequencies $\omega$. 
In terms of the coordinate $x$, the master equation takes the same wave equation form as the Schwarzschild electromagnetic perturbation equation. 
However, the parameter $c_{\text{eff}}$ modifies the tortoise coordinate, and therefore changes the effective potential when expressed as a function of $x$.
As will be seen more explicitly from the recurrence relation below, the resulting QNM spectrum is not obtained from the Schwarzschild electromagnetic one by a simple linear rescaling of the frequency.

To compute the spectrum, we employ Leaver's continued-fraction method \cite{Leaver:1985ax}. 
Although the QNM boundary conditions are most naturally stated in terms of the coordinate $x$, the Frobenius series is constructed in the radial coordinate $r$. 
In the following, we adopt the dimensionless normalization $2M\!=\!1$, and in the numerical analysis we restrict to the parameter region where $1+\ell-\ell^2/(2\kappa b^2)>0$, so that $c_{\rm eff}$ is real and the vector master equation is hyperbolic.
After extracting the ingoing and outgoing asymptotic behavior at the horizon and at infinity, respectively, we write
\begin{align}
u_3(r)=(r-1)^{-ik}\,r^{2ik}e^{ik(r-1)}
\sum_{n=0}^{\infty}a_n\left(\frac{r-1}{r}\right)^n\,,
\label{u3:ansatz}
\end{align}
where the parameter $k$ is related to the QNM frequency $\omega$ by
\begin{align}
k=\frac{\omega}{c_{\text{eff}}}\,.
\end{align}
Substituting the ansatz \eqref{u3:ansatz} into the radial equation yields the three-term recurrence relation
\begin{align}
c_0(n)a_n+c_1(n)a_{n-1}+c_2(n)a_{n-2}=0\,,\qquad n\ge1,
\end{align}
with $a_0=1$ and $a_{-1}=0$. The coefficients are given by
\begin{align}
c_0(n)&=n^2-2ikn\,,
\\
c_1(n)&=-2n^2\!+\!\left(8ik\!+\!2\right)n\!+\!8k^2\!-\!4ik\!-\!\frac{L(L\!+\!1)}{c_{\text{eff}}^2}\,,
\\
c_2(n)&=n^2-\left(4ik+2\right)n-4k^2+4ik\,.
\end{align}
An important point is that $c_{\text{eff}}$ does not enter only through the combination $k=\omega/c_{\text{eff}}$, but also appears explicitly in the angular term $L(L+1)/c_{\text{eff}}^2$. Therefore, the QNM frequencies are not related to the Schwarzschild electromagnetic ones by a simple linear scaling.

The QNM frequencies are determined numerically from the continued-fraction condition~\cite{Leaver:1985ax} associated with the above recurrence relation. In practical calculations, the infinite continued fraction is truncated at sufficiently large order. For highly damped modes, the convergence of the continued fraction is improved by using Nollert's asymptotic expansion for the tail $R_N$~\cite{Nollert:1993zz}. At large truncation order $N$, the tail takes the form
\begin{align}
R_N=-1\pm\sqrt{-2ik}\,N^{-1/2}+\left(2ik+\frac{3}{4}\right)N^{-1}+\cdots\,,
\end{align}
where the sign is chosen such that
\begin{align}
\text{Re}\!\left(\pm\sqrt{-2ik}\right)>0\,.
\end{align}
With this improvement, the continued-fraction method remains stable deep into the highly damped regime, allowing us to compute the overtone sequence shown in Fig.~\ref{fig1}(c) up to $n=400$, where every eighth overtone is displayed.

Figure \ref{fig1} summarizes the dependence of the QNM spectrum on $c_{\text{eff}}$ for $L\!=\!2$, using the convention $\omega=\omega_R-i\omega_I$, where $\omega_R$ denotes the oscillation frequency and $\omega_I>0$ characterizes the damping rate.
Panels (a) and (b) show that, as $c_{\text{eff}}$ increases, the real part of the frequency decreases, while the magnitude of the imaginary part increases. 
The modes therefore oscillate more slowly and decay more rapidly for larger $c_{\text{eff}}$. 
Panel (c) provides a global view of the same spectral deformation in the complex-frequency plane for $c_{\text{eff}}=0.8$, $0.9$, and $1.0$. 
The spectrum shifts systematically toward smaller $\omega_R$ and larger $\omega_I$ as $c_{\text{eff}}$ increases. 
This trend is visible not only for the lowest-lying modes but also along the overtone sequence, showing that the effect of the Lorentz-violating background persists beyond the fundamental modes.
Overall, although the master equation can be cast into a Maxwell-like form, the resulting QNM spectrum is not a trivial rescaling of the Schwarzschild electromagnetic one.
Instead, the effective speed $c_{\text{eff}}$ leaves a genuine imprint on the spectrum, lowering the oscillation frequency and increasing the damping rate throughout the QNM sequence.


\section{Discussion and conclusions}\label{Discussion}

In this work, we have carried out a systematic analysis of the full linear perturbations of the Schwarzschild-like black hole in Einstein-Bumblebee theory. 
By treating the metric and bumblebee field perturbations on the same footing, we have shown that the coupled gravitational-bumblebee system admits an exact hierarchical decoupling. 
At the level of the master equations, the perturbations separate into four decoupled wave equations. 
In each parity sector, one mode is governed by a Maxwell-like equation with an effective propagation speed $c_{\text{eff}}$, while the other obeys the standard Regge-Wheeler or Zerilli equation.
This decoupled structure leads to two key conclusions. 
First, the gravitational sector exhibits a dynamical immunity to the nonminimal Lorentz-violating coupling: although the bumblebee field enters the definition and reconstruction of the perturbation variables, the gravitational master equations retain the standard Schwarzschild Regge-Wheeler-Zerilli form. 
As a result, the odd- and even-parity gravitational modes remain isospectral. 
Second, the two bumblebee modes, one in the odd-parity sector and the other in the even-parity sector, obey the same Maxwell-like master equation and are therefore also strictly isospectral.
Thus, the full coupled system separates into a Schwarzschild-like gravitational sector and a Lorentz-violating Maxwell-like vector sector, each with its own odd-even isospectral structure.

Beyond the master equation structure, the reconstruction formulas provide an important consistency check on the degree-of-freedom content.
The reconstruction of the perturbations further clarifies how the propagating modes are encoded in the original metric and bumblebee variables. 
In the odd-parity sector, the metric perturbations are determined jointly by the gravitational and bumblebee master variables. 
In the even-parity sector, the reconstruction reveals an additional constrained quantity, $u_2(t,r)$, whose radial derivative is fixed by the master variables, while the function itself is determined only up to an integration function. 
This quantity therefore does not represent an additional propagating degree of freedom, but must be fixed by boundary, regularity, or other supplementary conditions.

We have also analyzed the QNMs of the Maxwell-like sector. 
The Lorentz-violating parameter enters the spectrum through the effective propagation speed $c_{\text{eff}}$ in a nontrivial way: the resulting QNM frequencies are not obtained from the Schwarzschild electromagnetic spectrum by a simple linear rescaling. 
Our numerical results show that increasing $c_{\text{eff}}$ lowers the oscillation frequency and enhances the damping rate, providing a characteristic spectral imprint of Lorentz-violating vector dynamics. 
From a phenomenological perspective, the mismatch between the standard gravitational propagation speed and the effective speed of the Lorentz-violating vector sector may lead to a relative delay between the corresponding ringdown signals. 
Although the bumblebee vector mode should not be identified directly with a standard electromagnetic counterpart, this effect provides a useful theoretical mechanism for connecting black-hole perturbations in Lorentz-violating gravity with multi-messenger tests. 
In this sense, $c_{\text{eff}}$ characterizes not only a deformation of the QNM spectrum, but also a potential timing observable for Lorentz-violating vector dynamics in multi-messenger black hole observations.

A further extension is to relax the fixed-norm branch assumed in this work. 
Let $X\!\equiv\! B^\mu B_\mu\pm b^2$. 
For a smooth quadratic potential $V(X)\!=\!\frac{1}{2}\lambda X^2$ \cite{Casana:2017jkc} with constant $\lambda$, $X\!=\!0$ defines the vacuum manifold, while fluctuations away from it correspond to a massive norm mode. 
Restricting to $\delta X\!=\!0$ can therefore be viewed as a consistent low-energy truncation in which this massive mode is frozen. 
In a Lagrange-multiplier realization $V(\lambda,X)\!=\!\frac{1}{2}\lambda X$ \cite{Maluf:2020kgf}, the same fixed-norm condition is imposed directly at the level of the action, so that $\delta X\!=\!0$ follows as part of the constraint rather than as a low-energy truncation.
Thus, the fixed-norm branch adopted here is well motivated in both descriptions, although the status of $\delta X\!=\!0$  is different in the two cases.
Relaxing this branch would activate the Higgs-like massive excitation associated with departures from the vacuum manifold \cite{Bluhm:2008yt}; the resulting perturbation theory would involve the coupled dynamics of metric perturbations, transverse/Nambu-Goldstone-like bumblebee modes \cite{Bluhm:2007bd}, and this massive mode, potentially leading to a modified decoupling structure and new spectral features. 
We leave this more general problem for future work.

Another natural direction is to ask whether the decoupling structure found here is a special property of vector-induced Lorentz violation, or a more general feature of spontaneously Lorentz-violating gravity. 
In this respect, gravity with a background Kalb-Ramond field provides a particularly interesting testing ground, since the Lorentz-violating order parameter is an antisymmetric tensor rather than a vector \cite{Kalb:1974yc,Kao:1996ea,Chakraborty:2014fva}. 
This  higher-rank structure may introduce new constraints and couplings in the perturbation equations, and it remains unclear whether a similar hierarchical decoupling, dynamical immunity of the gravitational sector, or odd-even isospectrality can survive.
Recent studies of Kalb-Ramond black holes have explored their solutions, shadows, QNMs, and other strong-field signatures \cite{Yang:2023wtu,Duan:2023gng,Liu:2024oas,Liu:2024lve,AraujoFilho:2025jcu,Lessa:2025kln,Guo:2023nkd,Shi:2025xkd,Xu:2025iwg,Yang:2025byw,Deng:2025atg,Gu:2025lyz,Liu:2025fxj,Xia:2025hwt,Xia:2025yzg,Al-Assi:2026jds,Sucu:2026nkw}. 
A fully coupled metric-Kalb-Ramond perturbation theory,  however, is still lacking. 
The method developed in this work offers a possible route toward such a theory: isolate the autonomous Lorentz-violating sector, reconstruct the constrained variables, and  test whether the induced gravitational sources can be absorbed into redefined master variables. 
This would allow a systematic comparison  between vector- and tensor-induced Lorentz violation in black hole spectroscopy and multi-messenger tests of gravity.


\section*{Acknowledgments}
This work is supported in part by the National Natural Science Foundation of China (Grants  No. 11875082, No. 12475056, No. 12547147), Gansu Province’s Top Leading Talent Support Plan, the Natural Science Foundation of Gansu Province (No. 22JR5RA389), and the 111 Project (Grant No. B20063), the China Postdoctoral Science Foundation (Grant No. 2025M783393).

\appendix
\begin{widetext}
\section{Original even-parity bumblebee equations}\label{AppA}
Before the rearrangement leading to Eqs. \eqref{eq:B1}-\eqref{eq:B3}, the linearized even-parity bumblebee equations contain explicit metric dependent source terms. 
In their original form, they read
\begin{align}
\begin{aligned}
{u''_0}{\!}-{\!}\dot{u}'_1+\frac{2}{r}\Big({u'_0}{\!}-{\!}\dot{u}_1\Big)-\frac{(1+\ell)(\mu^2{\!}+{\!}2)}{r^2f(r)}\Big({u_0}{\!}-{\!}{\dot{u}_2}\Big)=\frac{\ell\,b_r(r)}{\kappa b^2}\left[\dot{K}'-\frac{1-3f(r)}{2rf(r)}\dot{K}-\frac{1}{r}\dot{H}_2-\frac{\mu^2+2}{2r^2}H_1\right],
~~~~~~~~~~~~~~~~~~~~~~~~~~
\end{aligned}
\\
\begin{aligned}
\dot{u}'_0-\ddot{u}_1-\frac{\mu^2+2}{r^2}f(r)\Big({u_1}-{u'_2}\Big)=\,&\frac{\ell\,b_r(r)f(r)}{(\ell+1)\kappa b^2}\Bigg[f(r)K''+\frac{1+3f(r)}{2r}K'-\frac{f(r)}{2}H''_0-\frac{3f'(r)}{4}H'_0+\dot{H}'_1\\
&+\frac{f'(r)}{2f(r)}\dot{H}_1-\frac{1+3f(r)}{4r}H'_2\Bigg]
-\frac{\ell\,b_r(r)}{2\kappa b^2}\left[\ddot{H}_2+\frac{(\mu^2+2)f(r)}{r^2}H_2\right],~~~
\end{aligned}
\\
\begin{aligned}
u''_2\!-\!{u'_1}{\!}-{\!}\frac{1\!+\!\ell}{f(r)^2}\Big(\!\ddot{u}_2{\!}-{\!}\dot{u}_0\!\Big){\!}+{\!}\frac{f'(r)}{f(r)}\Big(u'_2{\!}-{\!}{u_1}\!\Big)+\frac{\ell^2f(r)^{-1}\!\!}{\kappa b^2r^2}u_2 = \frac{\ell\,b_r(r)}{2\kappa b^2f(r)}\Bigg[f(r)K'-f(r)H'_0-\frac{1-3f(r)}{2r}H_0+\dot{H}_1-\frac{1+f(r)}{2r}H_2\Bigg]\,.
\end{aligned}
\end{align}
\end{widetext}
%

\end{document}